\documentclass[twocolumn,twoside,slac_two]{revtex4}
\usepackage{epsfig}
\usepackage{graphicx}
\usepackage{fancyhdr}
\begin{document}
\voffset=-1.3cm

\title{Topics in Hadron Spectroscopy in 2009}

%

\author{Stephen Godfrey}
\affiliation{Ottawa-Carleton Institute for Physics,
Department of Physics, Carleton University, Ottawa, Canada K1S 5B6 \\
TRIUMF, 4004 Wesbrook Mall, Vancouver BC, Canada V6T 2A3}

\begin{abstract}
There have been numerous developments in hadron spectroscopy over the past year.  In 
this brief review I focus on two general areas.  In the first I give an update on
hadrons with $b$-quarks.  Their properties are in textbook agreement 
with QCD motivated constituent quark models. The second topic is the
Charmonium-like $XYZ$ states above open charm threshold that continue to be discovered.  
Many of these do not seem to be described
as conventional $c\bar{c}$ states and may be {\it hadronic molecules}, {\it Tetraquarks},
or {\it Charmonium Hybrids}.  The first topic is an example of how well we seem to understand
hadrons while the second case reminds us of how much we still have to learn.
\end{abstract}

\maketitle

\thispagestyle{fancy}

\section{Introduction}

There have been many developments in hadron spectroscopy during the past year. It is impossible
to do justice to all them in a brief review.  I will focus on two topics; new 
hadrons with $b$ quarks and the so called  charmonium $X$, $Y$, $Z$ states, many of which 
do not seem to be understood as conventional hadrons.  There are many other developments that
deserve attention but which I will not discuss:  the $f_{D_s}$ puzzle and possible hints for new
physics, 
the  $Y(2175)$ and $Y(10890)$  states, further measurements of $B_c$ 
properties, new measurements of hadronic transitions in quarkonium, 
recent results by the BESIII collaboration on $J/\psi$ decays, 
the $N^*$ program at Jefferson Lab, 
new measurements of quarkonium annihilation decays, etc. etc..  
I apologize for the many interesting topics I am not able to cover in a brief
update.  Some recent reviews of hadron spectroscopy are given in 
Ref.~\cite{Godfrey:2008nc,Eichten:2007qx,Harris:2008bv,Pakhlova:2008di,Olsen:2009ys}.

Quantum Chromodynamics (QCD) is the theory of the strong interactions but it has been a challenge
to calculate the properties of hadrons directly from the QCD Lagrangian in the regime
where the theory
is non-perturbative.  Instead, alternative approaches have been used; Lattice QCD, effective
field theories, chiral dynamics, and the constituent quark model.  Measurement of hadron 
properties provide an important test of these calculational approaches.  On the one hand, there
has been much progress in recent years, while on the other hand, a large number of states
have been discovered with properties that are not easily or consistently explained by theory.

In this context I use the constituent quark model (CQM) as a benchmark against which to identify 
and compare the properties of the newly discovered states \cite{Brambilla:2004wf}.  
Constituent quark models 
typically assume a QCD motivated potential that includes a Coulomb-like one-gluon-exchange
potential at small separation and a linearly confining potential at large separation.  The
potential is included in a Schrodinger or relativistic equation to solve for the eigenvalues
of radial and orbital angular momentum excitations.  For the case of mesons, the quantum numbers
are characterized by $J^{PC}$ quantum numbers where $S$ is the total spin of the quark-antiquark
system, $L$ is the orbital angular momentum, $P=(-1)^{L+1}$, and for self-conjugate mesons,
$C=(-1)^{L+S}$.  With these rules, the quark model predicts the allowed quantum numbers of
$J^{PC}=0^{-+}$, $1^{--}$, $1^{+-}$, $0^{++}$, $1^{++}$, $2^{++}\ldots$. Quantum
numbers not allowed by the CQM such as $J^{PC}=0^{--}$, $0^{+-}$, $1^{-+}$, and $2^{+-}$,
are often referred to as {\it exotic} combinations and, if such states were discovered, would
unambiguously signify hadronic states outside the quark model.  

In addition to the spin-independent potential there are spin-dependent potentials that
are relativistic corrections, typically assuming a Lorentz vector one-gluon-exchange and a 
Lorentz scalar confining potential. This leads to a short distance
spin-spin contact interaction which splits the spin-triplet and spin-singlet $S$-wave states. 
If the spin-spin interaction were not short range it would result in a splitting between
the spin-singlet and spin-triplet centre of gravity of the $L\neq 0$ states.
There is also a spin-spin tensor interaction which contributes to splittings in $S=1$, 
$L\neq 0$ multiplets in addition to mixings between states with the same $J^{PC}$ quantum
numbers.   Finally, there are spin-orbit interactions
that contribute to splittings between $S=1$, $L\neq 0$ states and mix states with unequal
quark and antiquark masses where $C$ is not a good quantum number and with the same $J^P$ such
as $^3P_1-^1P_1$ pairs.  The tensor and spin-orbit interactions give rise to the splittings
in, for example, the $\chi_{c0}, \chi_{c1}, \chi_{c2}$ multiplet.  
Strong Zweig allowed decays, annihilation decays, hadronic transitions, and electromagnetic
transitions have also been calculated using various models \cite{Brambilla:2004wf}.  
Putting all these predictions 
together one can build up a fairly complete picture of a quark model state's properties that can
be compared to experimental measurements.  

In addition to these conventional CQM hadrons, models of hadrons predict the existence
of additional states:
\begin{description}
\item[Hybrids] are states with an excited gluonic degree of freedom. 
Some hybrids are predicted to have exotic quantum numbers which would signal a non-$q\bar{q}$ state.   
Almost all models of hybrids predict that hybrids with conventional quantum numbers
will have very distinctive decay modes that can be used to distinguish them from conventional 
states.
\item[Multiquark States]
{\it Molecular States} are a loosely bound state of a pair of mesons near threshold.  One
signature of these states is that they exhibit large isospin violations.
{\it Tetraquarks} are tightly bound diquark-diantiquark states.  A prediction of
tetraquark models is that they are predicted to come in flavour multiplets.
\item[Threhold-effects] come about from rescattering near threshold due to the interactions 
between two outgoing mesons.  They result in mass shifts due to thresholds.  A related effect are
coupled channel effects that result in the mixing of two-meson states with $q\bar{q}$ 
resonances.
\end{description}
One can think of an analogy in atomic physics for multiquark states and hybrids. 
Say we know about atomic spectroscopy but theorists 
predict something they call molecules that have never been discovered.  Whether molecules
really exist would be an important test of theory.  Likewise, the unambiguous discovery of
hybrids and multiquark states is an important test of our models of QCD.

\section{Bottomonium $\eta_b$ State}

The observation of the $\eta_b$ is an important validation of lattice QCD and other calculations.  
One means of producing the $\eta_b$ is via radiative transitions, specifically M1 transitions
from the $n^3S_1(b\bar{b})$ states \cite{Godfrey:2001eb,Godfrey:2008zz}.  
The partial width for this transition is given by
\begin{equation}
\Gamma (^3S_1 \to ^1S_0 + \gamma)= \frac{4}{3} \alpha \frac{e_Q^2}{m_Q^2}
|\langle f | j_0(kr/2)| i\rangle |^2 k_\gamma^3
\end{equation}
where $e_Q$ is the quark charge in units of $e$, $m_Q$ is the mass of the quark, $k_\gamma$ is
the energy of the emitted photon, and $j_0$ is the spherical Bessel function.  
Hindered decays are those that occur between initial and final states
with different principle quantum numbers.  In the non-relativistic limit, 
the wavefunctions are 
orthogonal so that these decays are forbidden.  However, hindered decays
have large phase space so that even
small deviations from the non-relativistic 
limit can result in observable decays.  In contrast, the allowed 
transitions have very little phase space so the partial widths are likely to be
too small to be observed.

\begin{figure}[t]
\centering
\includegraphics[width=75mm, clip]{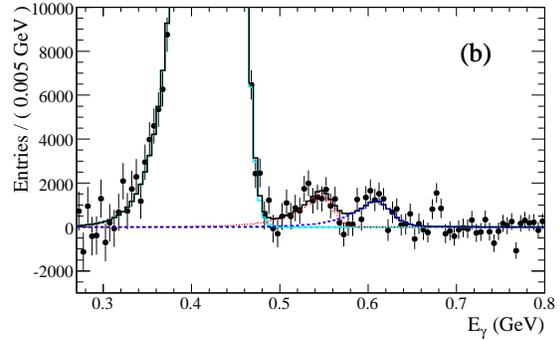}
\caption{From Babar Ref.~\cite{:2009pz}.
Photon spectrum for the $\Upsilon(2S)\to \eta_b \gamma$ transition after subtracting
the non-peaking background component, with PDFs for $\chi_{bJ}(2P)$ peak (light solid), 
ISR $\Upsilon(1S)$ (dot), $\eta_b$ signal (dash) and the sum of all three (dark solid). } 
\label{fig:babar_etab}
\end{figure}

Last year the BaBar collaboration announced the discovery of the $\eta_b(1^1S_0)$ state
in the transition $\Upsilon(3S)\to \eta_b \gamma$ \cite{:2008vj} with 
$B( \Upsilon(3S) \to \eta_b \gamma)=(4.8\pm 0.5 \pm 1.2)\times 10^{-4}$.  
More recently BaBar confirmed the $\eta_b$
in the transition $\Upsilon(2S)\to \eta_b \gamma$ 
with $B( \Upsilon(2S) \to \eta_b \gamma)=(4.2^{+1.1}_{-1.0} \pm 0.9)\times 10^{-4}$ \cite{:2009pz}.
The photon spectrum for this transition is shown in Fig~\ref{fig:babar_etab}.  
The average $\eta_b$ mass from the two measurements is 
\begin{equation}
M(\eta_b)=9390.4\pm 3.1 \hbox{ MeV}
\end{equation}
This is in agreement with Lattice QCD and the predictions of QCD based models.   
After the DPF conference, the CLEO collaboration \cite{Bonvicini:2009hs}
 reported evidence for the $\eta_b$ in 
$\Upsilon(3S)\to \eta_b \gamma$ with statistical significance of $\sim 4\sigma$ 
and with $M(\eta_b)=9391.8\pm 6.6 \pm 2.0$~MeV and
$B( \Upsilon(3S) \to \eta_b \gamma)=(7.1\pm 1.8 \pm 1.2)\times 10^{-4}$ which are consistent
with the BaBar measurements.
Both the measured mass and branching ratios support the models of heavy quarkonium spectroscopy.

\section{Bottomonium $\Upsilon(1D)$ State}

Another $b\bar{b}$ state I want to mention is the $\Upsilon(1D)$ state.  
It was suggested that
the $\Upsilon(1D)$ states could be observed in the cascade decays consisting of four
E1 transitions in the decay chain $3^3S_1 \to 2^3P_J \to 1^3D_J \to 1^3P_J \to 1^3S_1$ 
\cite{Godfrey:2001vc,Kwong:1988ae}.  
The E1 partial widths and branching ratios can be estimated using the quark model.
The CLEO collaboration followed this search strategy and observed an $\Upsilon(1D)$
\cite{Bonvicini:2004yj}. 
The data are dominated by the production of one $\Upsilon(1D)$ state consistent with the $J=2$
assignment.  It was measured to have a mass of $M=10161.1\pm 0.6 \pm 1.6$~MeV which
is in good agreement with predictions of potential models and Lattice QCD.
The measured BR for the decay chain is $B=(2.5\pm 0.5 \pm 0.5)\times 10^{-5}$ which compares well
to the predicted BR of $B=2.6\times 10^{-5}$.  

This result is not exactly a new result.  I mention it because CLEO's discovery 
was based on an $\Upsilon(3S)$ data sample of $5.8\times 10^6$ $\Upsilon(3S)$.  In contrast,
BaBar has collected a sample of $109\pm 1 \times 10^6$ $\Upsilon(3S)$'s, almost 20 times the size
of the CLEO sample.  BaBar has the potential to observe all three of the $1^3D_J$ states which
would be a nice test of our understanding of the $^3D_J$ splittings.

\section{Baryons with $b$ Quarks}

\begin{figure}[ht]
\centering
\includegraphics[width=38mm, clip]{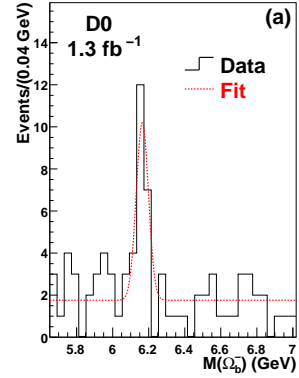} \\
\caption{ From D0 Ref.~\cite{Abazov:2008qm}.  
The $M(\Omega_b^-)$ distribution of the $\Omega_b^-$ candidates.  The
dotted curve is an unbinned lieklihood fit to the model of a constant background plus a 
Gaussian signal.   } 
\label{fig:d0omegab}
\end{figure}

\begin{figure}[ht]
\centering
\includegraphics[width=78mm, clip]{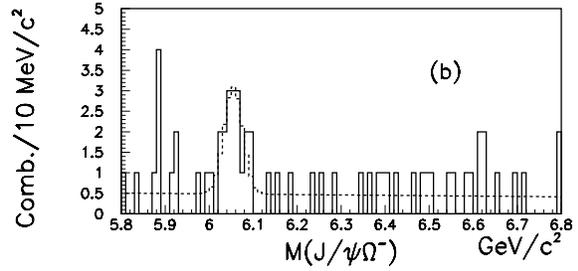} 
\caption{From CDF Ref.~\cite{Aaltonen:2009ny}. The invariant mass distribution 
of $J/\psi \Omega^-$.    } 
\label{fig:cdfomegab}
\end{figure}

\begin{figure}[hb]
\centering
\includegraphics[width=65mm, clip]{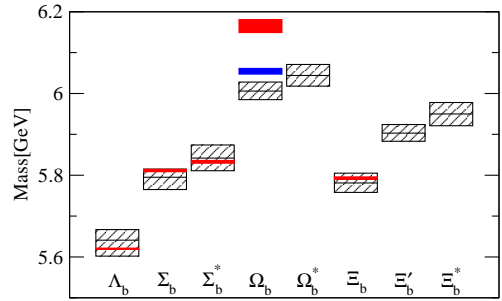} 
\caption{Masses of single-b baryons. The diagonally-hatched boxes are lattice results 
with combined statistical and systematic errors.  Solid bars (red) are experimental values
with the exception of the $\Omega_b$.  For the $\Omega_b$ the upper (red) bar is the D0 result and 
the lower (blue) bar is the CDF result. From Ref.~\cite{Lewis:2008fu} } 
\label{fig:hbmass}
\end{figure}

In the last year a number of baryons with $b$-quarks were observed for the first time by the 
D0 \cite{Abazov:2008qm} and CDF collaborations \cite{Aaltonen:2009ny}.  
The ground state baryons with $b$ quarks and their quark content are given by:
\begin{eqnarray}
& & \Lambda_b^0  =  |bud\rangle \cr
& & \Sigma_b^{(*)+}  =  |buu\rangle \cr
& & \Xi_b^{(*,\prime)-} =  |bsd\rangle \cr
& & \Omega_b^{(*)} = |bss\rangle 
\end{eqnarray}
The splittings of the ground state baryons can be described reasonably well by only including
the colour hyperfine interaction between two quarks \cite{Rosner:2006yk}
:
\begin{equation}
\Delta H_{ij}^{hyp}={{16\pi\alpha_s}\over{9m_i m_j}}
 \vec{S}_i\cdot\vec{S}_j \; \langle \delta^3(\vec{r}_{ij}) \rangle 
\sim  \gamma \; {{\vec{S}_i\cdot\vec{S}_j}\over {m_i m_j}}
\end{equation}
where we made the simplifying approximation that the wavefunction at the origin and $\alpha_s$
are roughly the same for all states.  This results in a number of predictions
\begin{equation}
M(\Sigma_b^*)-M(\Sigma_b)=[M(\Sigma_c^*)-M(\Sigma_c)] \times (\frac{m_c}{m_b})
\simeq 25\hbox{ MeV}
\end{equation}
\begin{eqnarray}
M(\Sigma_b)-M(\Lambda_b)& = & [M(\Sigma_c)-M(\Lambda_c)] \times \frac{(1-m_u/m_b)}{(1-m_u/m_c)} \cr
& \simeq & 192\hbox{ MeV}
\end{eqnarray}
to be compared to 
the measured splittings of  $21.2^{+2.0}_{-1.9}$~MeV and  192~MeV respectively which is 
very good agreement \cite{Karliner:2006ny}.  
While not all predictions are in such good agreement, this simple picture does 
work quite well.  A more careful
analysis by Karliner {\it et al.} \cite{Karliner:2008sv}
that includes wavefunction effects predicts
\begin{equation}
M(\Omega_b)= 6052.1 \pm 5.6 \hbox{ MeV}.
\end{equation}
This state was recently observed by the Fermilab D0 \cite{Abazov:2008qm} 
and CDF \cite{Aaltonen:2009ny}  collaborations in $J/\psi \Omega^-$.  The mass distributions
are shown in Fig.~\ref{fig:d0omegab} for D0 and in Fig.~\ref{fig:cdfomegab} for CDF 
with measured masses of
$M(\Omega_b)=6165 \pm 10  \pm 13 $~MeV and 
$M(\Omega_b)=6054.4 \pm 6.8 \;(stat) \; \pm 0.9 \; (sys)$~MeV by D0 and CDF 
respectively.   The two measurements are inconsistent.  
The CDF measurement is in good agreement with the quark model prediction and the lattice result 
while the D0 measurement is significantly larger.  The lattice results \cite{Lewis:2008fu}
along with observed ground state baryon masses 
with a $b$-quark are shown in Fig.~\ref{fig:hbmass}.

\section{The Charmonium like $X$, $Y$, $Z$ States}

\begin{figure}[t]
\centering
\includegraphics[width=90mm, clip]{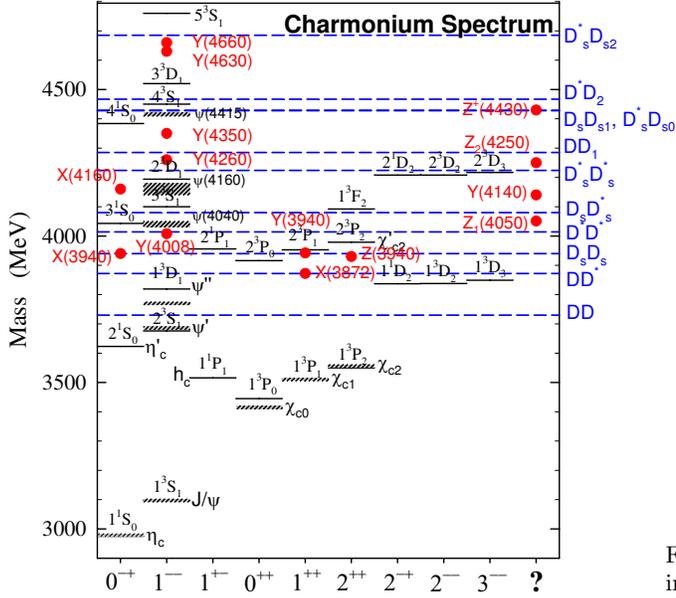} 
\caption{The Charmonium spectrum.  The solid lines are quark mode predictions 
\cite{Godfrey:1985xj}
the shaded lines are the observed conventional charmonium states \cite{Amsler:2008zzb},
the horizontal dashed lines represent various $D^{(*)}_{s} \bar{D}^{(*)}_{s}$ thresholds, 
and the (red) dots are 
the newly discovered charmonium-like states placed in the column with the 
most probable spin assignment.
The states in the last column do not fit elsewhere and appear to be truly exotic.} 
\label{fig:charmoniumb}
\end{figure}

Over the last five years or so, numerous charmonium like states have been discovered with many
of them not fitting into conventional charmonium spectroscopy
\cite{Godfrey:2008nc,Eichten:2007qx,Pakhlova:2008di,Olsen:2009ys}.  
This has led to considerable 
theoretical speculation that some of these new states are non-conventional hadrons like hybrids, 
molecules, or tetraquarks, or possibly some sort of threshold effect.  More and more of these 
states seem to appear every other day and it is far from clear what most of them actually are. The
charmonium spectrum is summarized in Fig.~\ref{fig:charmoniumb}.  This is a very cluttered figure
which underlines the complexity of the current situation.   A more 
detailed summary of these states is given in Table~\ref{tab:xyz_states}.

One can see that 
there are many of these charmonium like states.  I will restrict myself to the following;
I will start with the most recently observed states, the 
$Y(4140)$ seen by CDF \cite{Aaltonen:2009tz} and the $X(3915)$ seen by Belle \cite{Olsen:2009ys}.  
I will next report on 
the $Z^+$ states, charmonium-like states observed by Belle that carry charge so cannot be
conventional $c\bar{c}$ states.  I will then briefly discuss the $X(3872)$ which was the first
charmonium-like state to be observed and is the most robust, having been observed by many
experiments in different processes.  The final group is the $1^{--}$ 
$Y$ states observed in $e^+e^- \to \gamma_{ISR} +Y$.

\begin{table*}[t]
\caption{Summary of the Charmonium-like $XYZ$ states.}
\begin{center}
\begin{tabular}{lccclcll}
\hline\hline
\label{tab:xyz_states}
state    & $M$~(MeV) &$\Gamma$~(MeV)    & $J^{PC}$ & Seen In  & Observed by: & Comments \\ \hline
$Y_s(2175)$& $2175\pm8$&$ 58\pm26 $& $1^{--}$ & $(e^+e^-)_{ISR}, J/\psi \to Y_s(2175)\to\phi f_0(980)$ &  BaBar, BESII, Belle & \\
$X(3872)$& $3871.4\pm0.6$&$<2.3$& $1^{++}$ & $B\to KX(3872)\to \pi^+\pi^-
J/\psi$,$\gamma J/\psi$, $D\bar{D^*}$  &  Belle, CDF, D0, BaBar & Molecule?\\
$X(3915)$& $3914\pm4$& $28^{+12}_{-14}$ & $?^{++}$ & $\gamma\gamma\to \omega J/\psi$ & Belle & \\
$Z(3930)$& $3929\pm5$&$ 29\pm10 $& $2^{++}$ & $\gamma\gamma\to Z(3940)\to D\bar{D}$   &   Belle & $2^3P_2 (c\bar{c})$ \\
$X(3940)$& $3942\pm9$&$ 37\pm17 $& $0^{?+}$ & $e^+e^-\to J/\psi  X(3940)\to D\bar{D^*}$ (not
$D\bar{D}$ or $\omega J/\psi$) &   Belle & $3^1S_0 (c\bar{c})$? \\
$Y(3940)$& $3943\pm17$&$ 87\pm34 $&$?^{?+}$ & $B\to K Y(3940)\to \omega J/\psi$ (not
$D\bar{D^*}$)  & Belle, BaBar & $2^3P_1 (c\bar{c})$?\\
$Y(4008)$& $4008^{+82}_{-49}$&$ 226^{+97}_{-80}$ &$1^{--}$& $(e^+e^-)_{ISR}\to Y(4008)\to \pi^+\pi^- J/\psi$ &  Belle & \\
$Y(4140)$ & $4143\pm 3.1$ & $ 11.7^{+9.1}_{-6.2}$ & $?^{?}$ & $B\to K Y(4140) \to J/\psi \phi $ & CDF & \\
$X(4160)$& $4156\pm29$&$ 139^{+113}_{-65}$ &$0^{?+}$& $e^+e^- \to J/\psi X(4160)\to D^*\bar{D^*}$
 (not $D\bar{D}$) &   Belle & \\
$Y(4260)$& $4264\pm12$&$ 83\pm22$ &$1^{--}$&  $(e^+e^-)_{ISR}\to Y(4260) \to \pi^+\pi^- J/\psi$ & BaBar, CLEO, Belle & Hybrid?  \\
$Y(4350)$& $4324\pm24$&$ 172\pm33$ &$1^{--}$&  $(e^+e^-)_{ISR}\to Y(4350) \to \pi^+\pi^- \psi'$ & BaBar & \\ 
$Y(4350)$& $4361\pm13$&$ 74\pm18$ &$1^{--}$&  $(e^+e^-)_{ISR}\to Y(4350) \to \pi^+\pi^- \psi'$ &  Belle & \\
$Y(4630)$& $4634^{+9.4}_{-10.6}$ & $ 92^{+41}_{-32} $ &$1^{--}$&  $(e^+e^-)_{ISR}\to Y(4630)\to \Lambda_c^+\Lambda_c^-$ &  Belle    & \\
$Y(4660)$& $4664\pm12$&$ 48\pm15 $ &$1^{--}$&  $(e^+e^-)_{ISR}\to Y(4660)\to \pi^+\pi^- \psi'$ &  Belle    & \\
$Z_1(4050)$& $4051^{+24}_{-23}$&$ 82^{+51}_{-29}$ & ? &
$B\to K Z_1^{\pm}(4050)\to \pi^{\pm}\chi_{c1}$ &    Belle & \\
$Z_2(4250)$& $4248^{+185}_{-45}$&$ 177^{+320}_{-72}$ & ? &
$B\to K Z_2^{\pm}(4250)\to \pi^{\pm}\chi_{c1}$ &  Belle & \\
$Z(4430)$& $4433\pm5$&$ 45^{+35}_{-18}$ & ? & $B\to KZ^{\pm}(4430)\to \pi^{\pm}\psi'$ &   Belle & \\
$Y_b(10890)$  & $10,890\pm 3$ & $55\pm 9$ &   $1^{--}$ &
$e^+e^-\to Y_b\to \pi^+\pi^-\Upsilon(1,2,3S)$ &  Belle & \\
\hline\hline
\end{tabular}
\end{center}
\end{table*}

\subsection{The $Y(4140)$}

\begin{figure}[t]
\begin{center}
\centerline{\epsfig{file=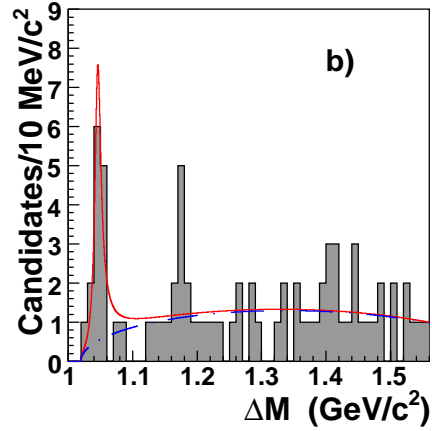,width=60mm,clip=}}
\end{center}
\caption{From CDF \cite{Aaltonen:2009tz}. 
Evidence for the $Y(4140)$ seen in $B\to J/\psi \phi$ with $J/\psi\to \mu^+\mu^-$ and
$\phi \to K^+K^-$.
The mass difference, $\Delta M$, between $\mu^+\mu^- K^+K^-$ and $\mu^+\mu^-$, in 
the $B^+$ mass window.  The dash-dotted (blue) curve is the background contribution and the solid (red)
curve is the total unbinned fit.  \label{cdf_y4140}}
\end{figure}

The CDF collaboration found evidence for the 
$Y(4140)$ in the $J/\psi \phi$ invariant mass distribution
from the decay $B^+\to J/\psi \phi K^+$ which is
shown in Fig.~\ref{cdf_y4140} \cite{Aaltonen:2009tz}.
The state has significance of $3.8\sigma$ with 
$M=4143.0 \pm 2.9 \pm 1.2$~MeV/c$^2$ and $\Gamma=11.7^{+8.3}_{-5.0}\pm 3.7$~MeV/c$^2$.  
Because both the $J/\psi$ and $\phi$ have $J^{PC}=1^{--}$ the $Y(4140)$ has +ve C parity.  Some 
argue that there are similarities to the $Y(3940)$ seen in $B\to J/\psi \omega K$ 
\cite{Abe:2004zs}.  

The question asked about all these new $XYZ$ states is: What is it?  
And as in all of these states, we consider the different
possibilities, comparing the state's properties to theoretical predictions.  
\begin{description}
\item[Conventional State]  The $Y(4140)$ is above open charm threshold so it would be expected
to have a large width which is in contradiction to its measured width.  
Hence, the $Y(4140)$ is unlikely to be a conventional $c\bar{c}$ state.
\item[$c\bar{c}s\bar{s}$ Tetraquark] A number of authors argue that the $Y(4140)$ is a 
tetraquark \cite{Mahajan:2009pj,Stancu:2009ka,Liu:2009ei}.  However a tetraquark
is expected to decay via rearrangement of the quarks with a width of $\sim 100$~MeV.  It is 
also generally expected to have similar widths to both hidden and open charm final states.  
The tetraquark interpretation does not, therefore, appear to be consistent with the data.
\item[Charmonium Hybrid]  Charmonium hybrid states are predicted to have masses in the 4.0 to 4.4 GeV 
mass range.  The $Y(4140)$'s mass lies in this range.  Hybrids are expected to decay 
predominantly to $SP$ meson pair final states with decays to $SS$ final state meson pairs
suppressed.  If the $Y(4140)$ were below $D^{**}D$ threshold the allowed decays to $D\bar{D}$ 
would be suppressed leading to a relatively narrow width.  The $D^*\bar{D}$ is an important mode
to look for.
\item[Rescattering via $D_sD^*_s$] Other possibilities are that the $Y(4140)$ is due to
$D_sD^*_s$ rescattering \cite{Rosner:2007mu} or the opening up of a new final state 
channel \cite{vanBeveren:2009dc,vanBeveren:2009jk}.
\item[$D^{*+}_s D^{*-}_s$ Molecule] The molecule explanation has been examined by a number of
authors \cite{Mahajan:2009pj,Branz:2009yt,Zhang:2009vs,Liu:2009ei,Albuquerque:2009ak,Ding:2009vd}.
The $D^{*+}_s D^{*-}_s$ threshold is $\sim 4225$~MeV implying a binding energy of $\sim 80$~MeV. 
If one interprets the $Y(3940)$ to be a $D^*\bar{D}^*$ molecule the binding energy of the
two systems are similar. Futhermore the decay $Y(4140)\to J/\psi \phi$ is similar
to the decay $Y(3940)\to J/\psi \omega$ although the widths are different.  The molecule picture
predicts that decays proceed via rescattering with decays to hidden and open charm final states 
equally probable. One should search for decays to open modes 
like $D\bar{D}$ and $D\bar{D}^*$.  Another prediction is that constituent mesons
can decay independently so observation of decays such as $Y(4140)\to D_s^{*+}D_s^-\gamma$ 
and $Y(4140)\to D_s^{+}D_s^{*-}\gamma$ would provide evidence for the molecule 
picture \cite{Liu:2009ei,Branz:2009yt,Liu:2009pu}.  
A $D^{*+}D_s^{*-}$ molecule is also predicted with mass 
$\sim 4040$~MeV with $J/\psi \rho $ as a prominent final state to look for\cite{Mahajan:2009pj}.

\end{description}
None of these explanations is compelling. 
A necessary first step to understand the $Y(4140)$ is to confirm it's existence in another
measurement as it has only been observed in one measurement at $3.8\sigma$.  It is then necessary
to observe other decay modes to  help to distinguish between the various possibilities.

\subsection{The $X(3915)$}

The $X(3915)$ is the most recent addition to the collection of $XYZ$ states (at least at the time 
of the conference). It was observed
by Belle in $\gamma\gamma\to \omega J/\psi$ with a statistical significance of
$7.5\sigma$ \cite{Olsen:2009ys}.  It has a measured mass and width of
$M=3914 \pm 3 \pm 2$~MeV and $\Gamma=23\pm 9 ^{+2}_{-3}$~MeV.  These parameters are consistent
with those of the $Y(3940)$.  The $2\gamma$ width times BR to $\omega J/\psi$ is 
$\Gamma_{\gamma\gamma} \times {\cal B}(X(3915)\to \omega J/\psi )= 69\pm 16^{+7}_{-18}$~eV 
assuming $J^P=0^+$ or $21\pm 4 ^{+2}_{-5}$~eV for $J^P=2^+$.  For comparison 
$\Gamma_{\gamma\gamma} \times {\cal B}(Z(3930)\to D\bar{D})= 180\pm 50 \pm 30$~eV.

\subsection{The $Z^+(4430)$, $Z_1^+(4050)$ and $Z_2^+(4250)$ States}

\begin{figure}[t]
\begin{center}
\centerline{\epsfig{file=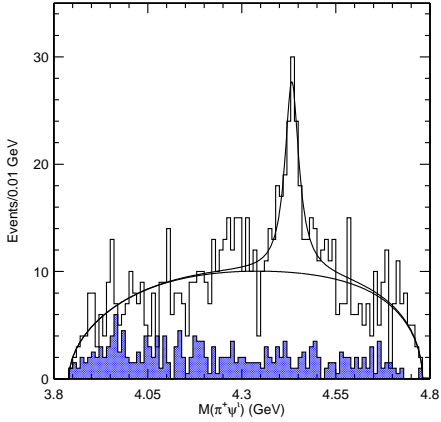,width=60mm,clip=}}
\end{center}
\caption{From Belle \cite{:2007wga}. The $M(\pi^+\psi')$ distribution. The shaded histogram show
the scaled results from the sideband region and the solid curves show the results of the fits.
\label{belle_z4430}}
\end{figure}

Belle observed a number of charmonium like states in $B$ decay that carry charge \cite{:2007wga},
 thus indicating that they cannot be conventional $c\bar{c}$ states.  The first state to be
discovered was the $Z^+(4430)$.  The  $\pi^+\psi (2S)$ 
invariant mass distribution is shown in Fig.~\ref{belle_z4430}.  The observed peak has a statistical
significance of $6.5\sigma$.  It's measured properties are $M=4433\pm 4\pm 2$~MeV,
$\Gamma=45^{+18}_{-12}\; ^{+30}_{-13}$~MeV, and 
${\cal B}(B^0\to K^{\mp}Z^{\pm})\times {\cal B} (Z^\pm\to \pi^\pm \psi')
=(4.1\pm 1.0\pm1.4)\times 10^{-5}$.  The unusual properties of the $Z^+(4430)$ led to the
usual explanations:
\begin{itemize}
\item $[cu][\bar{c}\bar{d}]$ Tetraquark \cite{Maiani:2007wz}
\item $D^*\bar{D}_1(2420)$ Threshold effect \cite{Rosner:2007mu}
\item $D^*\bar{D}_1(2420)$ $J^P=0^-, \; 1^-$ Molecule \cite{Meng:2007fu}.  The molecule 
explanation predicts that the $Z^+(4430)$ will decay into $D^*\bar{D}^*\pi$ and that 
it decays into $\psi(2S) \pi$ via rescattering.
\end{itemize}

\begin{figure}[t]
\begin{center}
\centerline{\epsfig{file=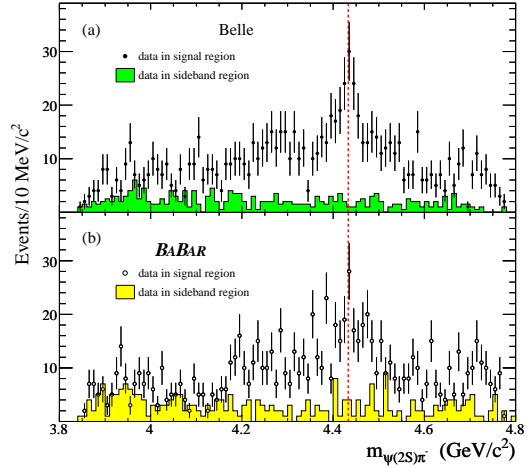,width=70mm,clip=}}
\end{center}
\caption{From BaBar Ref.~\cite{:2008nk}. (a) The $\psi(2S)\pi^-$ mass distribution from 
Ref.~\cite{:2007wga}; the data points represent the signal region and the shaded histogram
represents the background contribution. (b) shows the corresponding distribution form the BaBar
analysis.  The Dashed vertical line indicates $M_{\psi(2S)\pi^-}=4.433$~GeV/c$^2$.
\label{belle_babar1}}
\end{figure}

The Belle $Z^+(4430)$ observation was followed by a search by the BaBar 
collaboration in $B\to K \pi^\pm \psi(2S)$ \cite{:2008nk}.  BaBar performed a detailed 
analysis of the $K\pi^-$ system, corrected for efficiency, and included $S$, $P$, and $D$ waves 
in their analysis.  Fig.~\ref{belle_babar1} shows the invariant mass distributions from Belle and
Babar \cite{:2008nk}.  While there appears to be an excess of events in the $Z^+(4430)$ region
in the BaBar data, BaBar finds no conclusive evidence in their data for the $Z^+(4430)$.

More recently Belle performed a complete Dalitz analysis \cite{:2009da}.  Belle confirms the
 $Z^+(4430)$ with $M=4443^{+15+17}_{-12-13}$~MeV and $\Gamma = 109^{+86+57}_{-43-52}$~MeV.
The width is larger than the original measurements but the uncertainties are large.

\begin{figure}[t]
\begin{center}
\centerline{\epsfig{file=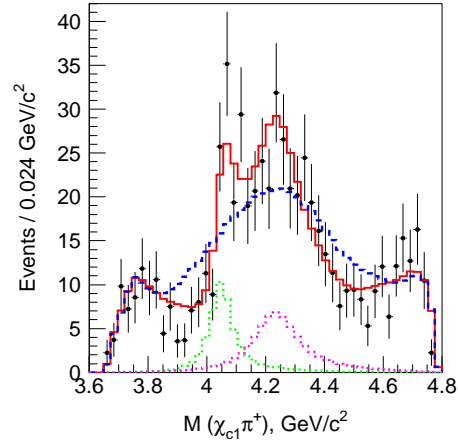,width=60mm,clip=}}
\end{center}
\caption{From Belle Ref.~\cite{Mizuk:2008me}. The $M(\chi_{c1\pi^+})$ distribution for the 
Dalitz plot slice $1.0\hbox{ GeV}^2/c^4 < M^2(K^-\pi^+) < 1.75\hbox{ GeV}^2/c^4$. The
dots with error bars represent data, the solid (dashed) histogram is the Dalitz plot fit 
result for the fit model with all known $K^*$ and two (without any)  $\chi_{c1}\pi^+$ 
resonance, the dotted histograms represent the contribution of the two $\chi_{c1}\pi^+$ 
resonances. 
\label{belle_zs}}
\end{figure}

The Belle collaboration has also observed two resonance structures in $\pi^+ \chi_{c1}$ mass 
distributions shown in Fig.~\ref{belle_zs}  \cite{Mizuk:2008me} with masses
and widths of $M_1=4051 \pm 14 ^{+20}_{-41}$~MeV, $\Gamma_1=82^{+21+47}_{-17-22}$~MeV
and  $M_2=4248 ^{+44+180}_{-29-35}$~MeV, $\Gamma_2=177^{+54+316}_{-39-61}$~MeV.

Belle has now found evidence for three charged charmonium like objects.  If confirmed they
represent clear evidence for some sort of multiquark state, either a molecule or tetraquark.  
Confirmation is needed for all three of them.

\subsection{The $X(3872)$}

The $X(3872)$ is probably the most robust of all the charmonium like objects.  It was first
observed by Belle as a peak in $\pi^+\pi^-J/\psi$ in
$B^+\to K^+ \pi^+\pi^-J/\psi$ \cite{Choi:2003ue}.
It was subsequently confirmed by CDF \cite{Acosta:2003zx}, 
D0 \cite{Abazov:2004kp}, and BaBar \cite{Aubert:2004ns}.  The PDG \cite{Amsler:2008zzb}
values for its mass and width are $M=3872.2\pm 0.8$~MeV and $\Gamma=3.0^{+2.1}_{-1.7}$~MeV.
Unlike most other $XYZ$ states there is a fair amount known about the $X(3872)$ properties. 
The radiative transition $X(3872)\to \gamma J/\psi$ has been observed by Belle \cite{Abe:2005ix}
and by BaBar \cite{Aubert:2006aj} and more recently $X(3872)\to \psi(2S) \gamma$ by 
BaBar \cite{:2008rn}.  This implies that the $X(3872)$ has $C=+$.  A study of
angular distributions by Belle favours $J^{PC}=1^{++}$ \cite{Abe:2005iya} while a higher
statistics study by CDF allows $J^{PC}=1^{++}$ or $2^{-+}$ \cite{Abulencia:2006ma}.
In the decay $X(3872)\to \pi^+\pi^-J/\psi$ the dipion invariant mass is consistent
with originating from $\rho\to \pi^+\pi^-$ \cite{Abulencia:2006jp}.  
The decay $c\bar{c}\to \rho J/\psi$ violates isospin and should be strongly suppressed.
The decay $X(3872)\to D^0\bar{D}^0\pi^0$ has been seen by Belle \cite{Gokhroo:2006bt}
and the decay $X(3872)\to D^0\bar{D}^0 \gamma$ by BaBar \cite{Aubert:2007rva}.  These decays imply
that the $X(3872)$ decays predominantly via $D^0\bar{D}^{*0}$.  To understand the nature of the
$X(3872)$ we work through the now familiar possibilities and compare the theoretical predictions 
for each case to the $X(3872)$ properties.

\subsubsection{Conventional Charmonium}

These possibilities were discussed in 
Ref.~\cite{Barnes:2003vb,Eichten:2004uh,Barnes:2005pb,Eichten:2005ga}.  The 
$1^1D_2$ and the $2^3P_1$ are the only conventional states with the correct quantum numbers
that are close enough in mass to be associated with the $X(3872)$.  However, both these 
possibilities have problems.  Another new state, the $Z(3921)$, is identified with the $2^3P_2$ 
state implying that the $2P$ mass is $\sim 3940$~MeV.  Identifying the $X(3872)$ with the 
$2^3P_1$ implies a spin splitting much larger than would be expected. If the $X$ were the 
$1^1D_2(c\bar{c})$, the radiative
transition $1^1D_2 \to \gamma 1^3S_1$ would be a highly suppressed M2 transition so that the
observation of $X(3872)\to \gamma J/\psi$ disfavours identifying the $X(3872)$ as 
the $1^1D_2$ state.

\subsubsection{Tetraquark}

This possibility was proposed in Ref.~\cite{Maiani:2004vq}.  This scenario predicts more nearly
degenerate states including charged states which have yet to be observed. A high statistics 
study by CDF of the $X(3872)$ mass and width tested the hypothesis of two states and finds
$\Delta m < 3.6 (95\% C.L.)$ with $M=3871.61\pm 0.16 \pm 0.19$~MeV \cite{Aaltonen:2009vj}.  
The mass splitting of the 
$X(3872)$ states 
produced in charged and neutral $B$ decays is consistent with zero. These measurements
disfavour the tetraquark interpretation.

\subsubsection{$D^0 D^{*0}$ Molecule}

The molecule explanation appears to be the most likely interpretion of the $X(3872)$
\cite{Close:2003sg,Voloshin:2003nt,Swanson:2003tb,Braaten:2005ai}.  It is very close to the 
$D^0D^{*0}$ threshold so it is quite reasonable that it is 
an $S$-wave bound state.  One of the early predictions of the molecule interpretation 
is that \cite{Swanson:2003tb}:
\begin{equation}
\Gamma(X(3872)\to \rho J/\psi) \simeq \Gamma(X(3872)\to \omega J/\psi)
\end{equation}
so that large isospin violations are expected.  On the other hand, the
decays $X(3872)\to \gamma J/\psi$ and $X(3872)\to \gamma \psi(2S)$ \cite{:2008rn}
indicate it has $c\bar{c}$ content.  The most likely explanation is that both the $X(3872)$ 
and $Y(3940)$ have 
more complicated structure, consisting of mixing with both $2^3P_1(c\bar{c})$ and
$D^0D^{*0}$ components 
\cite{Godfrey:2006pd,Danilkin:2009hr,Ortega:2009hj,Matheus:2009vq,Kalashnikova:2009gt}.  
This may also explain the unexpected large partial width for 
$Y(3940)\to J/\psi \omega$ \cite{Aubert:2007vj}.

\subsection{$Y$ States in ISR ($J^{PC}=1^{--}$)}

There are now six ``$Y$'' states seen in $e^+e^-\to \gamma_{ISR} Y$.  Because they are seen
in ISR they have $J^{PC}=1^{--}$.
Because of time and space constraints I will only discuss
two of these states; the $Y(4630)$ observed by Belle \cite{Pakhlova:2008vn} 
which is one of the newest
$Y$ states  and the $Y(4260)$ first observed by BaBar \cite{Aubert:2005rm} 
which is one of the oldest.

\subsubsection{$Y(4630)$}

\begin{figure}[t]
\begin{center}
\centerline{\epsfig{file=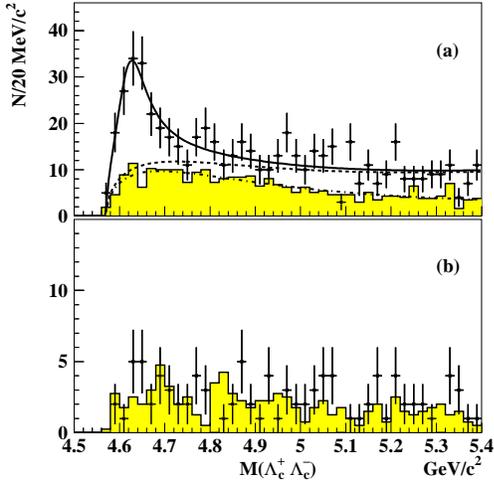,width=65mm,clip=}}
\end{center}
\caption{From Belle Ref.~\cite{Pakhlova:2008vn}. The $M_{\Lambda^+_c\Lambda^-_c}$ spectrum.
(a) With $\bar{p}$ tag. The solid curve represents the result of the fit, the threshold function
is shown by the dashed curve, and the combinatorial background parametrization is shown by the
dashed-dotted curve.  (b) With  proton (wrong-sign) tag.  Histograms show the normalized 
contributions from $\Lambda_c^+$ sidebands.
\label{belle_y4630}}
\end{figure}

The $Y(4630)$ was seen by the Belle collaboration in 
$e^+e^-\to \Lambda_c^+\Lambda_c^- \gamma_{ISR}$  with mass and width
$M=4634 ^{+8+5}_{-7-8}$ and $\Gamma=92^{+40+10}_{-24-21}$ \cite{Pakhlova:2008vn}.  The 
$\Lambda_c^+\Lambda_c^-$ mass distribution is shown in Fig~\ref{belle_y4630}.
There are some speculations about what it might be.  A possible explanation 
is that it is a dibaryon threshold effect.  A similar effect is also seen by Belle in 
$B\to \Lambda_c^+ \bar{p} \pi^-$ \cite{Abe:2004sr} with a $6.2\sigma$ peak observed at
threshold in the $\Lambda_c^+ \bar{p} $ invariant mass distribution.  Other possibilities
put forward are to identify the the $Y(4630)$ with the $Y(4660)$, also observed in ISR,
but in the $\pi^+\pi^-\psi'$ final state, or to identify the $Y(4630)$ with the $5^3S_1$ 
charmonium state.

\subsubsection{$Y(4260)$}

\begin{figure}[t]
\begin{center}
\centerline{\epsfig{file=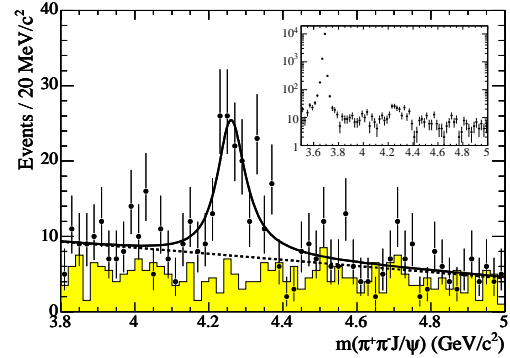,width=68mm,clip=}}
\end{center}
\caption{From BaBar Ref.~\cite{Aubert:2005rm}. The $\pi^+\pi^- J/\psi$ invariant mass spectrum
in the range 3.8-5.0~GeV/c$^2$ and (inset) over a wider range that includes the $\psi(2S)$. 
The points with error bars represent the selected data and the shaded histogram represents
the scaled data from neighbouring $e^+e^-$ and $\mu^+\mu^-$ mass regions.  The solid curve
shows the result of the single-resonance fit and the dashed curve represents
the background component.
\label{babar_y4260}}
\end{figure}

The $Y(4260)$ was the first of the $Y$ states to be observed. It was first observed
by the BaBar collaboration as an enhancement in the $\pi\pi J/\psi$ final state in
$e^+e^- \to \gamma_{ISR} J/\psi \pi\pi$ \cite{Aubert:2005rm}.
The $\pi^+\pi^- J/\psi$ invariant mass
distribution is shown in Fig.\ref{babar_y4260}  \cite{Aubert:2005rm}.
BaBar found further evidence for the $Y(4260)$ in $B\to K (\pi^+\pi^- J/\psi)$ 
\cite{Aubert:2005zh} and it was also independently confirmed by
CLEO  \cite{Coan:2006rv}  and Belle \cite{:2007sj}.  Thus,  it is the oldest and most robust
of the $Y$ states.  The possibilities for the $Y(4260)$ are:
\begin{description}
\item[Conventional Charmonium] The first unaccounted $1^{--}$ state is the $\psi(3D)$ 
with predicted mass $M[\psi(3D)]\sim 4500$~MeV which is much heavier than the observed
mass.  Thus, the $Y(4260)$ appears to represent an overpopulation of the expected $1^{--}$
states.  In addition, the absence of open charm production speaks against it being 
a conventional $c\bar{c}$ state.  There was the suggestion that the $Y(4260)$ could be
identified as the $\psi(4S)$ state \cite{LlanesEstrada:2005hz}, displacing the $\psi(4415)$
from that slot although the authors acknowledge this fit is somewhat forced.
\item[Tetraquark] Maiani {\it et al.}, \cite{Maiani:2005pe} proposed that the $Y(4260)$
is the first radial excitation of the $[cs][\bar{c}\bar{s}]$.  They predict that the
$Y(4260)$ should decay predominantly to $D_s \bar{D}_s$ and predict a full nonet of 
related four-quark states.
\item[$D_1D^*$ Bound State] Close and Downum \cite{Close:2009ag}
suggest that two $S$-wave mesons can be bound via pion exchange leading to a spectroscopy
of quasi-molecular states above 4~GeV and a possible explanation of the $Y(4260)$ and
$Y(4360)$.  They suggest searches in $D\bar{D}3\pi$ channels as well as in $B$ decays.
\item[$c\bar{c}$ Hybrid] This has been suggested in a number of papers 
\cite{Zhu:2005hp,Close:2005iz,Kou:2005gt}. This possibility has a number of attractive features.
 The flux tube model \cite{Isgur:1984bm} and lattice QCD \cite{Lacock:1996ny}
predict the lowest $c\bar{c}$ hybrid at $\sim 4200$~MeV.  LGT suggests searching for other 
closed charm models with $J^{PC}=1^{--}$ such as $J/\psi \eta$, $J/\psi \eta'$, 
$\chi_{cJ} \omega \ldots$ \cite{McNeile:2002az}.  
Most models predict that the lowest mass hybrid mesons will
decay to $S+P$-wave mesons final states \cite{Kokoski:1985is,Close:1994hc}.  
The dominant decay mode is expected to be $D^+D_1(2420)$.  The $D_1(2420)$ has a width of
$\sim 300$~MeV to $D^*\pi$ which suggests to search for the $Y(4260)$ in $DD^*\pi$ final states.
Evidence of a large $DD_1(2420)$ signal would be strong evidence for the hybrid interpretation.
Note that searches for these decays by Belle find no evidence \cite{Pakhlova:2007fq}. 
Another prediction of the hyrid explanation is to search for partner states.  The flux tube model
predicts a multiplet of states nearby in mass
with conventional quantum numbers; $0^{-+}$, $1^{+-}$, $2^{-+}$,
$1^{++}$, $1^{--}$ and states with {\it exotic} quantum numbers $0^{+-}$, $1^{-+}$, $2^{+-}$.
Identifying some of these $J^{PC}$ partners would further validate the hybrid scenario.
\end{description}

\subsubsection{$Y$ States in ISR:  What are they?}

There are now six $Y$ states observed in ISR.  I've described the possibilities 
for the $Y(4260)$ but the same process of elimination follows for all of them.  The
measured $Y$ masses don't match the peaks in the $D^{(*)}\bar{D}^{(*)}$ cross sections and 
there does not appear to be room for additional conventional $c\bar{c}$ states in this mass 
region unless the predictions are way off.  
It has been suggested that many of the $Y$-states are multiquark
states, either tetraquarks or molecules. 
Molecules are generally believed to lie just below threshold
and are bound via $S$-wave rescattering and pion exchange.  Few of the $Y$-states lie close 
to thresholds so at best this might explain special cases but cannot be a general explanation.
Other problems with the multiquark explanation are discussed below.
The final possibility considered is that some of the $Y$ states are charmonium hybrids.  The 
$Y(4260)$ is the most robust of all these states and is quite possibly a hybrid. Most of the
$Y$-states, however, need confirmation and  
more detailed measurements of their properties.

\section{Summary}

During the past year there have been many new developments in hadron spectroscopy. 
In some cases the new results reinforce our understanding in the context 
of the constituent quark model
while in other cases they demonstrate that we still have much to learn.

Many hadrons with heavy quarks have been observed and their properties are in good agreement 
with theory.  The observation of the $\eta_b$ by BaBar in the electromagnetic transitions
$\Upsilon(3S)\to \gamma \eta_b$ and $\Upsilon(2S)\to \gamma \eta_b$ provides further 
evidence that QCD motivated quark models and Lattice QCD calculatons
are essentially correct.  Likewise, the properties
of the ground state baryons with $b$ quarks are well described by the simplest of quark 
model assumptions to the point that they can be used as a  homework problem in a particle
physics course.

In contrast, it is not at all clear what most of the new charmonium-like $XYZ$ states 
are. 
There are now something like 16 charmonium like $XYZ$ states with new ones, seemingly, 
discovered every other day.  
A few can be identified as conventional states and a few more,
the $X(3872)$ and $Y(4260)$ for example, are strong candidates for hadronic molecule
and hybrid states.
These latter two are the best understood, having been confirmed by several experiments and
observed in different processes and channels.

It has been suggested that many of the $XYZ$ states are multiquark
states, either tetraquarks or molecules. The problem with the tetraquark explanation is that it
predicts multiplets with other charge states that have not been observed, and 
larger widths than have been observed.  The possibility that some of the $XYZ$ states
are molecules is likely intertwined with threshold effects that occur when channels are opened up.
Including
coupled-channel effects and the rescattering of charmed meson pairs in the mix can also
result in shifts
of the masses of $c\bar{c}$ states and result in meson-meson binding which could help explain the 
observed spectrum \cite{Voloshin:2006pz,Close:2009ag,Danilkin:2009hr,vanBeveren:2009fb}.  
In my view, a comprehensive study including coupled channels is a necessity if we are to 
understand the charmonium spectrum above $D\bar{D}$ threshold.  

Many of the $XYZ$ states need independent confirmation and to understand them will require 
detailed studies of their properties. 
With better
experimental and theoretical understanding of these states we will have more confidence 
in believing that any of these new states are non-conventional $c\bar{c}$ states
like molecules, tetraquarks, and hybrids.

Hadron spectroscopy continues to intrigue with a bright future.
There is the potential for many new measurements; 
BaBar has considerable unanalyzed data that might hold evidence for new states,  Belle and
BESIII have bright futures, and JLab, PANDA, and the LHC promise to produce exciting new physics
in the longer term.

\begin{acknowledgments}
This work was supported in part by the Natural Sciences and Engineering Council of Canada.
\end{acknowledgments}


\end{document}